# Virtual ALMA Tour in VRChat: A Whole New Experience

Masaaki Hiramatsu (National Astronomical Observatory of Japan),
S_Asagiri (Virtual Space Program), Stella. G. Amano (Virtual Space Program),
Naohiro Takanashi (The University of Tokyo), Shio K. Kawagoe (The University of Tokyo),
Kazuhisa Kamegai (National Astronomical Observatory of Japan)

**Abstract**

Many forefront observatories are located in remote areas and are difficult to visit, and the global pandemic made visits even harder. Several virtual tours have been executed on YouTube or Facebook Live, however, it is difficult to feel a sense of immersion and these are far from the actual experience of visiting a site. To solve this problem, we pursued an astronomy outreach event on the virtual reality social platform VRChat. To provide an experience similar to visiting the site, we performed a virtual tour of the ALMA Observatory in VRChat guided by an ALMA staff member. 47 guests participated in the tour. The post-event survey showed that the overall lecture and guided tour were very positively accepted by the participants. Respondents answered that the communication in the VRChat was more intensive than in other online outreach events or on-site public talks. The ratio of respondents who answered that they were able to communicate well with the guide was higher for those who used head mounted displays than for those who participated in other ways. 40 answered that the tour increased their interest in astronomy, and this did not show a clear difference depending on how they participated. In the free descriptions in the responses, there were noticeable mentions of the physical sensations received from the realistic 3D space, which left a positive and strong impression on the participants. The responses show that VRChat has the potential to be a strong tool for astronomy communication in the pandemic and post-pandemic eras.

## Introduction

Astronomy observatories in various locations of the world and in orbit are producing a number of amazing pictures of the universe and unravelling a variety of cosmic mysteries. The results are disseminated daily through websites, press releases, and social media. People can access the latest results through these media, however, they face difficulties learning about the research process.

Visiting the observatories and getting explanations from the researchers and engineers actually working there is one important way to understand how astronomy research is carried out. As such, many observatories accept public visits[1]. In Chile, there are numerous large-scale astronomical observatories, and visits to astronomical facilities and other Astro-tourism projects are highlighted by the national government[2]. Other locations such as Hawaii, the Canary Islands, and China are also active in Astro-tourism with astronomical observation facilities at the core.

One drawback of visiting astronomical observatories is that most of the cutting-edge facilities are in remote areas to avoid light pollution and radio interference from cities and to seek a clear atmosphere. This makes facility visits time consuming and sometimes require special safety considerations. An extreme example is the summit facility of the Atacama Large Millimeter/submillimeter Array (ALMA)[3], which is located at an altitude of 5000 metres, so visits by the general public are restricted to prevent altitude sickness. These and other hurdles make a visit to observatories a very limited experience and inaccessible to many people. The global pandemic of Covid-19 has raised these hurdles even higher. The pandemic forced heavy restrictions on both international and domestic travels, and many observatories suspended their public visit programmes as a precaution against infection. This resulted in a significant loss of opportunities for the non-specialist public to come into contact with the places where cutting-edge astronomy research is being carried out.

---

[1] Selected examples of the observatory public visits are the Atacama Large Millimeter/submillimeter Array (https://www.almaobservatory.org/en/outreach/alma-observatory-public-visits/), European Southern Observatory's Paranal Observatory (https://www.eso.org/public/about-eso/visitors/paranal/), and W. M. Keck Observatory (https://keckobservatory.org/visit/).

[2] The governmental agency for tourism in Chile highlights Astro-tourism for visitors on its official website, including information about tours of astronomical facilities: https://chile.travel/en/what-to-do/astrotourism

One way to overcome the geographical difficulty is live webcast from observatories. One pioneering example of the virtual observatory tour is "Around the World in 80 Telescopes" held during the International Year of Astronomy 2009 (Pierce-Price et al., 2009). The webcast connected 80 observatories around the world one after another over a 24-hour period to introduce the facilities and astronomical highlights. It gave the public an unprecedented experience to visit such a large number of observatories and to see life at observatories.

In the pandemic era, several observatories have provided online virtual tours. For example, the European Southern Observatory (ESO) organised virtual tours to their Paranal and La Silla Observatories[4]. They use omnidirectional images of the observatories as a staff member explains how the observatories look, how telescopes are operated, and how the observatory staff works. The tour is streamed on their Facebook page and YouTube channel. Questions from viewers are submitted via chat and answered by the guide. This form of virtual tour can provide some sense of realism and two-way communication through questions and answers between the viewers and the guide. However, even using omnidirectional photos, it is difficult to feel immersed in the enormity of actual telescopes in a flat video window.

By wearing a head-mounted display (HMD), immersive virtual reality (VR) allows one to enter a world created by three-dimensional (3D) computer graphics (CG). In this report we use the word "immersive" to refer to the feeling of the experiencers as if they were present in the virtual world. There are several examples using VR for astronomy communication. Arcand et al. (2018) describes the visualisation of the supernova remnant Cassiopeia A so that people can virtually walk through the object and understand its structure. Ferrand and Warren (2018) describes a VR demo in an open house of a research institute. Visitors wore a headset and viewed volume renderings of actual science data of supernovae and supernova remnants while receiving explanations from a researcher acting as a guide. Kersting et al. (2021) reported a VR tour featuring gravitational astronomy at a science festival and discussed the effectiveness of the immersive VR environment for engagement in education and public outreach activities.

---

[3] ALMA is a radio astronomy observatory located in the Atacama desert, northern Chile. It consists of 66 parabolic antennas combined to function as one giant telescope.

[4] Announcement about virtual guided tours from the European Southern Observatory:
https://www.eso.org/public/announcements/ann20020/

The ALMA Project of the National Astronomical Observatory of Japan (NAOJ) has created an immersive virtual reality tour of the ALMA facilities in 2016. The tour consists of a 4.5-minute compilation of the omnidirectional photos and fulldome videos of the observatory with the recorded voiceover, and viewers can watch the video with a HMD. NAOJ ALMA Project has used this VR content at the NAOJ Open House and the VR video is now available on YouTube[5].

At the Open House, visitors could watch the video through a GearVR headset or a smartphone in an HMD set-up. After they watched the video, they commented that VR video provided a good sense of immersion and most people indicated that they were very satisfied with the realism of the experience. The comments well described one of the profound affordances of VR, the feeling of presence (Johnson-Glenberg, 2018). However, the viewpoint is limited to a fixed location in the video. It is impossible to walk through the video, and there is no live interaction with the researchers during watching the video.

VR is also being used for outreach in fields other than astronomy. Tibaldi et al. (2020) reported a VR program featuring the 3D immersive visualisation of volcanic outcrops, which used both for scientific research and public outreach. Gochman et al. (2019) developed a VR tool to simulate the sight of tarsiers. They conducted demos in schools and received very positive feedbacks for better understanding of visual optics and evolution. Having VR contents in public events and schools can create real-time communication, unlike a VR experience with a pre-recorded video described earlier. However, in a pandemic, it is difficult for people to come together and share a single device, so this virtual experience would only be safe in certain circumstances.

In this article, we report a guided tour of the virtual ALMA observatory created through the VR social platform VRChat in response to the Covid-19 pandemic and discuss the possibility of breaking through the difficulties such as:

- in visiting remote observatories,
- in having the feeling of presence with the traditional virtual tours using YouTube or facebook live,
- in organising VR tours in an on-site science event in the pandemic era.

## Virtual ALMA Observatory Tour
### Virtual ALMA Observatory in VRChat

VRChat[6] is a platform where people use avatars to communicate with other users in 3D CG worlds through voice and gestures. In VRChat, any user can virtually construct various "worlds" and other users can visit them. One can join the VRChat worlds either through an HMD connected with a PC or regular flat display using VRChat's Desktop mode. In the Desktop mode, although the sense of immersion and the freedom of gesture are limited, users can still communicate with their voice and move freely around the world as one can with an HMD. One of the features of VRChat is that participants do not need to gather in one place or share a device, as they can connect via the Internet[7] from home or elsewhere. This feature can be an advantage to organising events under the pandemic. A precedent similar to VRChat is Second Life and Gauthier (2007) describes an example of astronomy communication on Second Life.

---

[5] This VR tour video is available on the YouTube ALMA Japan Channel with Japanese narration:
https://www.youtube.com/watch?v=e3T_p4MPlWo

[6] VRChat (https://hello.vrchat.com/) is compatible with most of HMDs connected with a PC, but not compatible with stand-alone-type HMDs, PlayStation VR, and VR using smartphone.

[7] As for the internet bandwidth, while 10 Mbps may be enough for conversations and motion tracking on VRChat, loading the world data could be a bottleneck. For example, the size of the ALMA world is 62.24MB, so at 10Mbps, it takes about 50 seconds to load the data. The recommended bandwidth for stress-free participation

As of May 2019, the number of unique users in VRChat was 430,000[8], and the user base showed a large growth during the pandemic[9]. As of May 2019, 75% of users were under 35 years old and 43% of users were under 25 years old, which means that the user base is heavily skewed towards the younger generation. Although gender distribution is highly biased: 81.5% of the users are male, it is not addressed in the context of this work.

Two of the authors of this report are members of the Virtual Space Program (VSP)[10], a group of people who are interested in astronomy and space development and are active mainly in VRChat (RORERU et al., 2020). VSP has produced several virtual worlds in VRChat, such as a planetarium, museum of space probes and space telescopes, and a rocket launch pad, and organises virtual tours regularly. VSP made a presentation introducing their activity at the annual meeting of the Japanese Society for Education and Popularization of Astronomy held in August 2020. After this presentation, the other authors of this report contacted with the VSP members to discuss possible future collaboration. In the discussion the authors came up with an idea to build a virtual ALMA world and organise a virtual tour of the world.

After the initial discussion for the collaboration, S_Asagiri of VSP voluntarily made a virtual world of the ALMA Observatory in VRChat (Figure 1). The world consists of 66 realistic antenna models and terrain. The real-scale antenna models were created based on photos of the actual ALMA antennas and partly based on the blueprint of miniature model of the Japanese ALMA antennas made by NAOJ. Four types of the ALMA antenna models were created using Blender, a free and open-source 3D-CG software toolset. The created models were exported to Unity, which is a cross-platform game engine, via fbx format. Unity is the only toolkit to upload 3D models to VRChat. The colors and textures of the antennas were adjusted on Unity. It took about 10-20 hours to model each type of the ALMA antenna, one hour to create the terrain, and a few hours to place each antenna to the actual configuration. The antenna location is based on the actual ALMA antenna configuration in January 2016[11], which was displayed in real-time on the observatory's website at that time. The antenna array was in its compact configuration [12]. In the compact configuration, the antennas are densely packed together, and it is easy to see the whole group of antennas. Hiramatsu, as the East Asian ALMA Education and Public Outreach Officer at NAOJ, offered advice on how to make the world more real, including the colour of the sky and the ground, the terrain, the shape of the antenna pads, and detailed shape of the antennas. Input from a person with in-depth knowledge of the real

---

is 100Mbps.

[8] Wagner James Au, "VRChat Site User Demographics: 430,000 Uniques, Mostly Male And Over 25", New World Notes blog: https://nwn.blogs.com/nwn/2019/05/vrchat-user-numbers-demographics-social-vr.html

[9] Ben Lang, "Social VR App 'VRChat' is Seeing Record Usage Amidst the Pandemic", Road to VR: https://www.roadtovr.com/vrchat-record-users-coronavirus/

[10] The website of Virtual Space Program (VSP) (Japanese): https://virtualspaceprogram.org/

[11] The ALMA antenna configuration is available in this short video: https://www.youtube.com/watch?v=-UcrXSs39U0

[12] The ALMA Cycle 3 Technical Handbook describes the antenna configuration: https://almascience.nao.ac.jp/documents-and-tools/cycle3/alma-technical-handbook

telescopes and first-hand experience of the site greatly improved the reality of the virtual world.

The visitors to the virtual ALMA Observatory can move freely around the antennas as if they are on the actual site. One feature unique to VRChat that cannot be realized in the real world is that one moves through the air and observe the antenna array from above. In addition, one can move the virtual antennas synchronously by pressing a button in the VR world. These features help participants better understand how the ALMA antennas work as a single telescope. The VR world also has miniature models of the four types of the ALMA antennas and users can pick up the antenna model and take a closer look. These miniature models are helpful to find out differences in the designs of the antennas developed by Japan, US, and Europe.

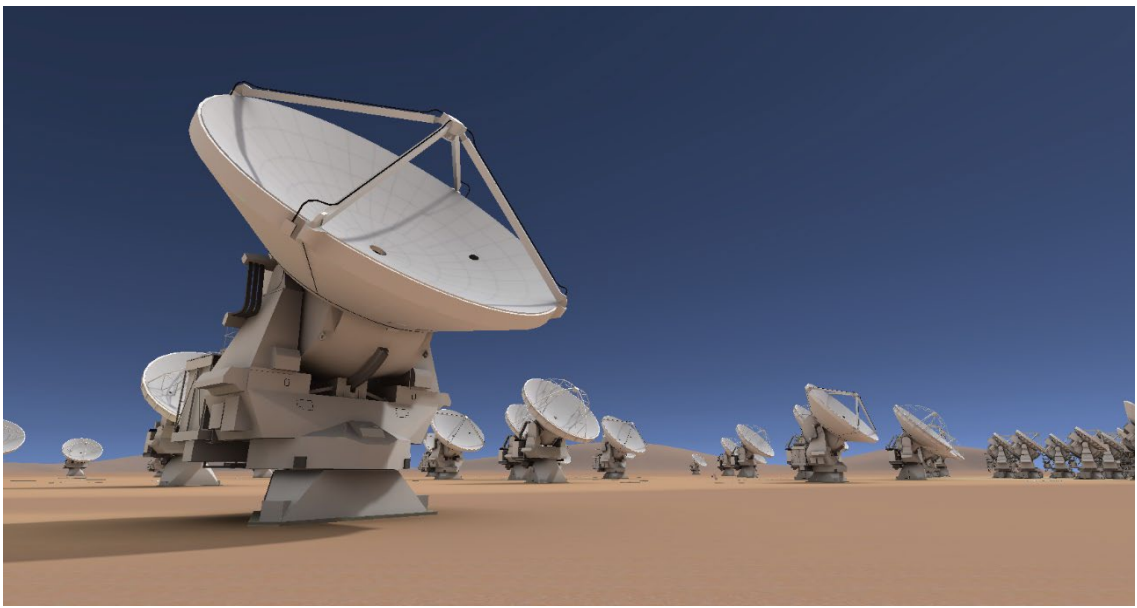

*Figure 1.* Virtual ALMA Observatory made in VRChat. Credit: VSP

Prior to the virtual tour, a preview was held with a limited number of people to improve the ALMA world and to plan the tour. This test was held because some of the organisers of the virtual ALMA tour had never experienced the VRChat. The testers were selected from the TENPLA project (Takanashi & Hiramatsu 2018), a group of astronomers and communicators, including the authors of this report. With this test the organisers were able to familiarize themselves with the world of VRChat and also conceptualize and test the content of the virtual tour. In addition, this test was an opportunity to improve the reproducibility of the virtual ALMA world, such as the colour of the sky and terrain, and the position of the Sun and the mountains. After the virtual tour, this ALMA world opened to the public in VRChat and anyone can access it now. There is a simple virtual explanation panel written in Japanese in the ALMA world and visitors can freely walk around the world. The cumulative number of the visitors to the world is 2025 as of September 5, 2021.

### Setting the Tour

The virtual tour of the ALMA Observatory in VRChat was held in the evening (18:00-19:00 local time) of Saturday 17 October 2020, free of charge. We selected this timeslot to maximize the number of participants. This event was advertised with VSP's Twitter and Discord, NAOJ ALMA Project's Twitter, and an article on a news website featuring VR. The upper limit of the participant in a VRChat World is set to 80, however, the maximum number of people who can comfortably experience a VRChat world depends on the amount of data in the world and the avatars used by the participants. To avoid the system overload, we had to limit the number of participants. As a result, the actual number of total participants in the VRChat was 47, and one guide. To further reduce the load on the servers, VSP distributed lightweight avatar data and asked the participants to use the avatar. In addition to the tour in VRChat, the tour was streamed on YouTube[13]. One of the tour organisers from VSP held a virtual camera in the VRChat and its output was broadcast to YouTube. Participants via YouTube watched this streaming video, and therefore they could not change the angle of view of the video by themselves and participate in the conversation taking place in the VRChat. Table 1 summarizes what the participants can do through HMD, desktop mode, and YouTube. The highest simultaneous YouTube viewership of the tour was 61 people and a total of 159 people joined the stream by the end of the tour. The entire virtual tour lasted about an hour and ten minutes and after that some participants and the guide stayed behind for additional questions and explanations. Since most of the participants stayed until the end of the event, the organisers consider that one hour was about the right length for the event. If it had been longer, we could have given a more detailed introduction to the technical features and scientific achievements of ALMA, but as a first tour, we thought it was important to give an overview and let the participants enjoy the world by looking around the virtual observatory, rather than deepening the knowledge of ALMA. For participants who want to know more details, it would be better to provide another opportunity.

| Action | HMD | Desktop Mode | YouTube |
|---|---|---|---|
| **Walk freely** | Yes | Yes | No |
| **Voice Communication** | Yes | Yes | No |
| **Gesticulate** | Yes (HMD controller tracks body motion) | Yes (use a keyboard to gesture) | No |
| **Chat via text** | No | No | Yes (texts cannot be seen from VRChat) |

*Table 1. Summary of what the participants can do through HMD, desktop mode, and YouTube.*

Table 2 shows the tour programme. The tour started with a 30-minute introductory presentation of ALMA in the middle of the virtual ALMA world (Figure 2). Hiramatsu was the lecturer and guide for the tour. He made a presentation file with Microsoft PowerPoint and sent it to the VSP organiser so that the presentation was shown in the VRChat world. In this presentation, the lecturer explained why astronomers observe radio waves from celestial objects, why multiwavelength astronomy is important, why ALMA was built in the Atacama Desert in Chile, as well as ALMA's representative results on planet formation, distant galaxies, and the discovery of complex organic molecules around young stars.

---

[13] The recording of the virtual ALMA tour on VRChat is available on YouTube: https://www.youtube.com/watch?v=NYCsUA0r_Y0

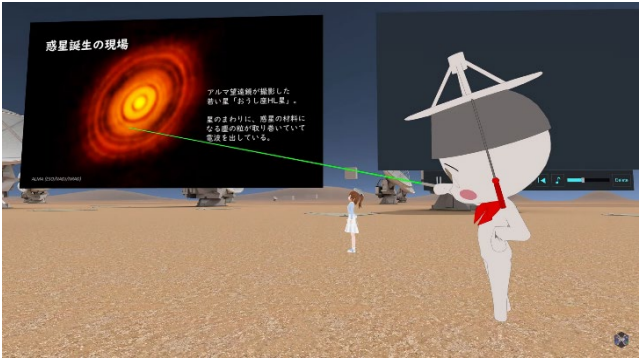

*Figure 2. Overview lecture of ALMA. PowerPoint presentation is embedded in the virtual world. The lecturer wears an avatar inspired by an ALMA antenna. Credit: VSP*

| Time (JST) | Content |
|---|---|
| 18:00-18:30 | Introductory lecture of ALMA, explaining significance of multiwavelength astronomical observations, and introducing ALMA's representative results such as the protoplanetary disk around a young star HL Tau and the most distant detection of oxygen at a galaxy 13.28 billion light years away. |
| 18:30-18:50 | Virtual guide in the ALMA world, explaining the basic function of the ALMA antennas, technological differences in the antennas developed by Japan, US, and Europe, and actual operation of the telescope including the reconfigurations of the antennas. |
| 18:50-19:10 | Questions and Answers session at the middle of the antenna array. |

*Table 2. The timetable of the virtual tour*

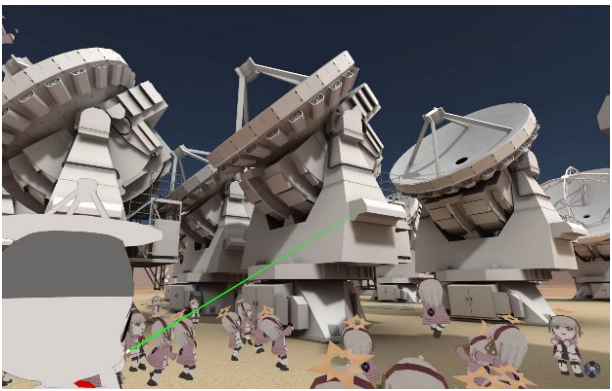

*Figure 3. Masaaki Hiramatsu, the East Asian ALMA Education and Public Outreach Officer, using his avatar representing an antenna (bottom left) explains ALMA's 7-meter antenna array developed by Japan in the VRChat CG world. In order to avoid overloading the server with individual participants using different heavy-weight avatars, all participants use the same avatar. Credit: VSP*

After the presentation, the lecturer guided the participants to the vicinity of the virtual array to explain the technical details of the antennas, including design ingenuity to achieve sufficient accuracy, differences in the antennas manufactured by Japan, US, and Europe, and how the antenna configuration is changed (Figure 3). During the tour, participants in VRChat were able to look around the antenna from any angle. They also freely express their impressions through voice chat as the lecturer explained.

An interesting point is that they were not merely reacting alone, but listening to the voices of other participants around them and responding to them. In contrast to the virtual tours via Facebook or YouTube that allow people to express their impressions via text chat, VRChat allows communication via voice. The participants were able to have two-way communication more easily and naturally in a similar environment to a real tour.

At the end of the tour, we had a question and answer session. The participants surrounded the lecturer and asked questions (Figure. 4). A wide variety of questions were asked, such as the process to make a radio image from observational data, the cost to construct ALMA, radio interference, and data calibration after antenna relocation. Questions about the instruments from a technical point of view were raised during the tour and those were apparently evoked from watching the realistic antenna models, but the questions that came up during the Q and A session were more general in nature and not much different from those in typical face-to-face lectures.

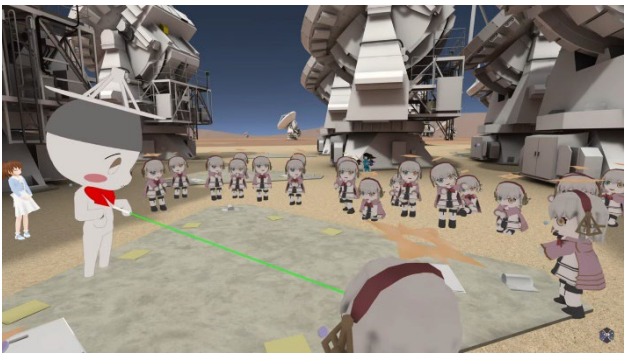

*Figure 4. Questions and Answers session at the end of the tour. This session was held in the middle of the virtual antenna array. Credit: VSP*

## Results of Questionnaire to the Participants

At the conclusion of the tour, we asked the participants to answer the questionnaire on their experiences via Google Forms[14]. We received 44 responses. In this section, we summarise the responses and discuss the advantages of using VRChat in astronomy communication.

### User Environments and their Background

Among 44 respondents, 21 were in their 20s or younger, accounting for nearly half of all participants[15]. This agrees with the overall demographics of VRChat but is in contrast to the audience for the NAOJ public lecture featuring ALMA held in a face-to-face format in Tokyo in February 2020, in which only 15% of the participants (35 out of 239) are in their 20s or younger. VRChat is a good way to approach and communicate with the younger generation with a high interest in the cutting-edge technology of VR.

We asked why they participated in the tour by presenting three options (1. Because it is a VSP event, 2. because it is in the virtual ALMA world, and 3. because an expert will give a talk) and they were allowed to select multiple options. Out of 44, 43 selected the option 3. It is still uncommon for a professional astronomer to give a talk on VRChat, and it was the first time for VSP to invite an astronomer to their event. Several participants highly evaluated in the questionnaire responses that the organisers invited an expert to give a talk in the VR world, which shows that there is a high demand for expert's participation. 34 and 27 selected option 1 and 2, respectively. 16 out of 44 did not know about ALMA before the event. Considering that more than 75% of the participants chose the option 1, it can be said that most of the participants are already familiar with VR and are interested in advanced technologies like VR.

We asked about the level of interest in astronomy before the event. Figure 5 shows graphs of the responses from VSP members and others separately. In total, 25 answered that they were "very interested" in astronomy, 16 answered "some level of interest", and 3 had "little interest". The proportion of the responses was almost the same between VSP members and others. Since the event was mainly advertised via VSP's and NAOJ ALMA Project's social media, it is natural that most of the participants are already interested in astronomy. To increase the participation of people who are not interested in astronomy, we would strengthen announcements on web media that cover general VR topics.

---

[14] The survey was done in the Japanese language. Quotations in these sections are literal translations.

[15] Organisers of VR events should be aware that there are restrictions on the use of HMDs for people under certain age (around 12 years old).

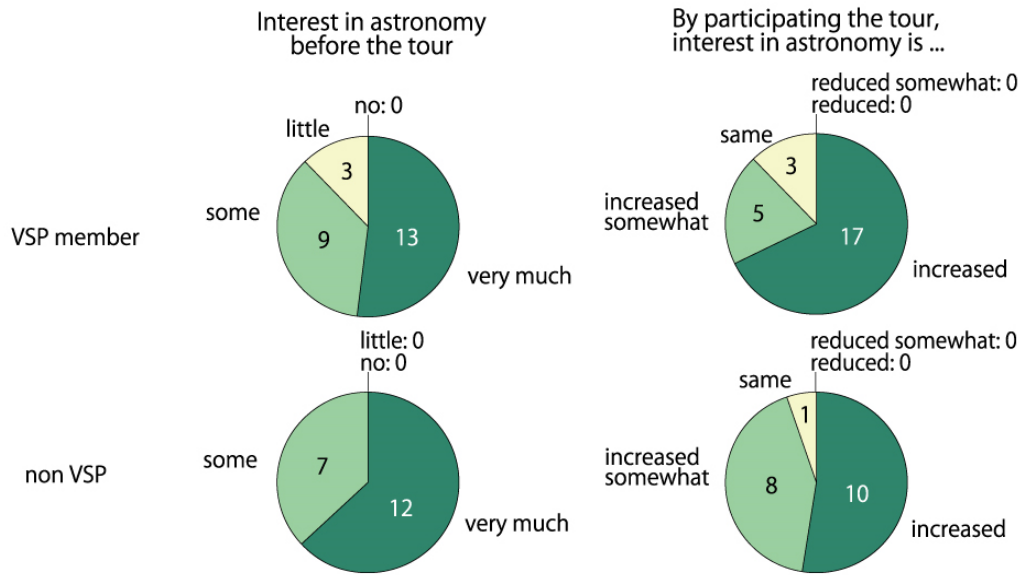

*Figure 5.* Summary of survey responses. The diagrams show the difference in responses depending on the environment through which the respondent participated in the lecture by the participating environment. Credit: Masaaki Hiramatsu

We asked if their interest in astronomy had increased after this event. 27 said it had, 13 said it had increased somewhat, and 4 said it had remained the same. Of the 41 participants who said they were "very interested" or "somewhat interested" in astronomy before participation, 38 said the interest was "increased" or "somewhat increased" after the tour. Among the three who answered that they had "little interest" in astronomy before the event, two answered that the interest somewhat increased after the event, and one answered that the interest level keeps same. This result indicates that the program had the effect of arousing further interest in astronomy in participants who were already interested, and also had a positive impact on the participants less interested in astronomy.

We asked the participants if they had attended any in-person or online science public lectures prior to this tour. Out of 44, 27 answered that they had attended in-person science lectures before, and 16 answered that they had attended online lectures. Half of those who have attended an in-person lecture said they have never attended an online lecture. Considering that the demand for expert's talks as described above, this indicates that there is still a great potential for VR lectures, but that the science side has not yet captured it.

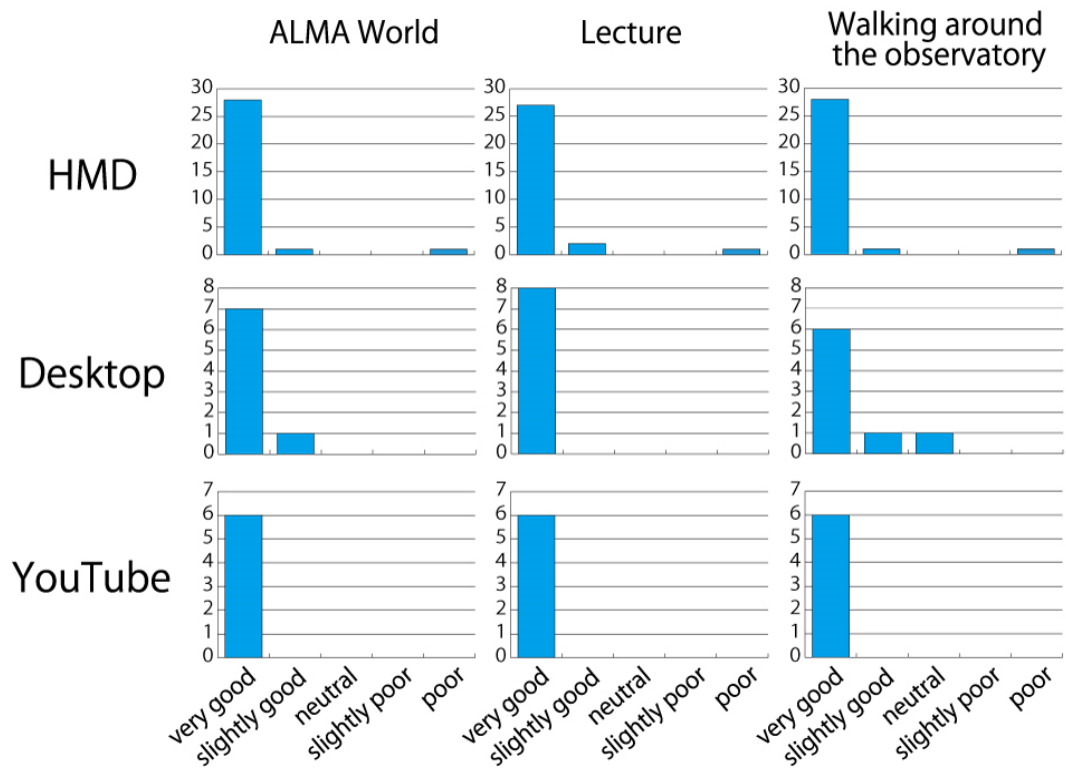

*Figure 6.* Survey responses on the satisfactory level for the ALMA world, lecture, and walking around the virtual observatory. Credit: Masaaki Hiramatsu

**User Experience with Different Media**

Among 44 respondents, 30 joined the tour with an HMD, 8 joined with the Desktop mode of VRChat, and 6 watched the streaming on YouTube. There was no significant difference in the overall satisfaction level with the tour based on the medium through which they participated (HMD, Desktop mode, or YouTube) as shown in Figure 6.

A possible reason for this is that the participants with the Desktop mode and YouTube have less experience of HMDs and do not know how the experience with HMD was. In the free-text feedback from participants who used HMDs, there were many positive comments attributed to the immersive experience of using HMDs, as described in a later section, so it is possible that the experience when using HMDs is more positive than when the same person uses other methods. In order to have a full-scale experience using HMDs, however, users need to prepare their own equipment, but this requires a certain level of cost (more than 2000 US dollars for an HMD and a powerful PC) and it is not easy to attract a wide range of participants if we organise an event only in VRChat. The fact that this tour received high evaluations regardless of the participants' environment shows that a hybrid event combining the use of HMD and YouTube streaming can also have a high outreach effect. One respondent answered "poor" to all the questions, but wrote comments in the free text section praising the event highly, which might indicate that he/she mistakenly selected the answers.

To see the difference in user experience through different media, we separately summarised the responses in Figure 7. More than 70% of the HMD users answered that they had interacted with the lecturer a lot or to some extent. This shows that the immersive environment possibly makes spontaneous communication easier. The free-write responses described in the following sections and personal feeling by the guide supports this interpretation. Although the numbers of responses for Desktop mode and YouTube were small and it is difficult to be definitive, the fraction of those who had good interaction with the guide was smaller than those with HMD. Users who participated through the Desktop mode could use their voice and gestures with their avatars. Although it was not an immersive environment, the survey responses show that some levels of communication took place even with the Desktop mode (survey responses indicate that this occurred but no further detail on what kind of communication was described). In contrast, we did not have a way to communicate directly with YouTube viewers. The participants through YouTube can use text chat to interact with other YouTube viewers, but there were no channels of communication with other participants and the guide using HMDs and the Desktop mode because they could not see the YouTube chat. The two respondents from YouTube who answered "very much" to this question may highly value one-way communication, rather than two-way communication.

Next, we discuss the communication between the participants. Judging from the survey responses, the participants with HMDs had slightly better communication compared with those with Desktop mode and YouTube, however, not many participants communicate with other participants. This weak communication among participants would be due to the overall structure of the event. A lecture and a guided tour are mainly based on communication between the lecturer/guide and each participant, and communication between participants is outside the main scope of this programme. Within VRChat, it is also possible to organise activities by dividing participants into small groups to interact, since voices are heard louder when they are close together in the virtual space and quieter when they are farther away, just like in a real space. If organisers are able to invite more than two experts, they could organise a virtual observatory tour in smaller groups that would better facilitate communication among participants and between participants and guides.

## Difference between Face-to-Face Events and VRChat

To find out how participants felt about the event in the VRChat world, we asked those who had attended face-to-face lectures in the past about the biggest difference between the face-to-face event and this virtual tour. 27 out of 44 respondents answered that they have attended in-person lectures, and 24 described the differences. Two characteristics in VR emerged from the responses. One was the realistic experience, and the other was the closeness to the lecturer.

The HMDs provide an immersive experience, allowing one to look around the realistic virtual antenna like they could do if they were actually at the observatory. Selected responses to this question are as follows.
- "It was as if I had actually been there in Chile."
- "The explanation with the virtual model in front of me is easier to experience it as something that is happening to me proactively than the video or slides, and it seems to attract more interest."
- "Since the lecture was given in the VR world, which is a reproduction of the observatory site, it left a stronger impression than explanations and questions using only slides."

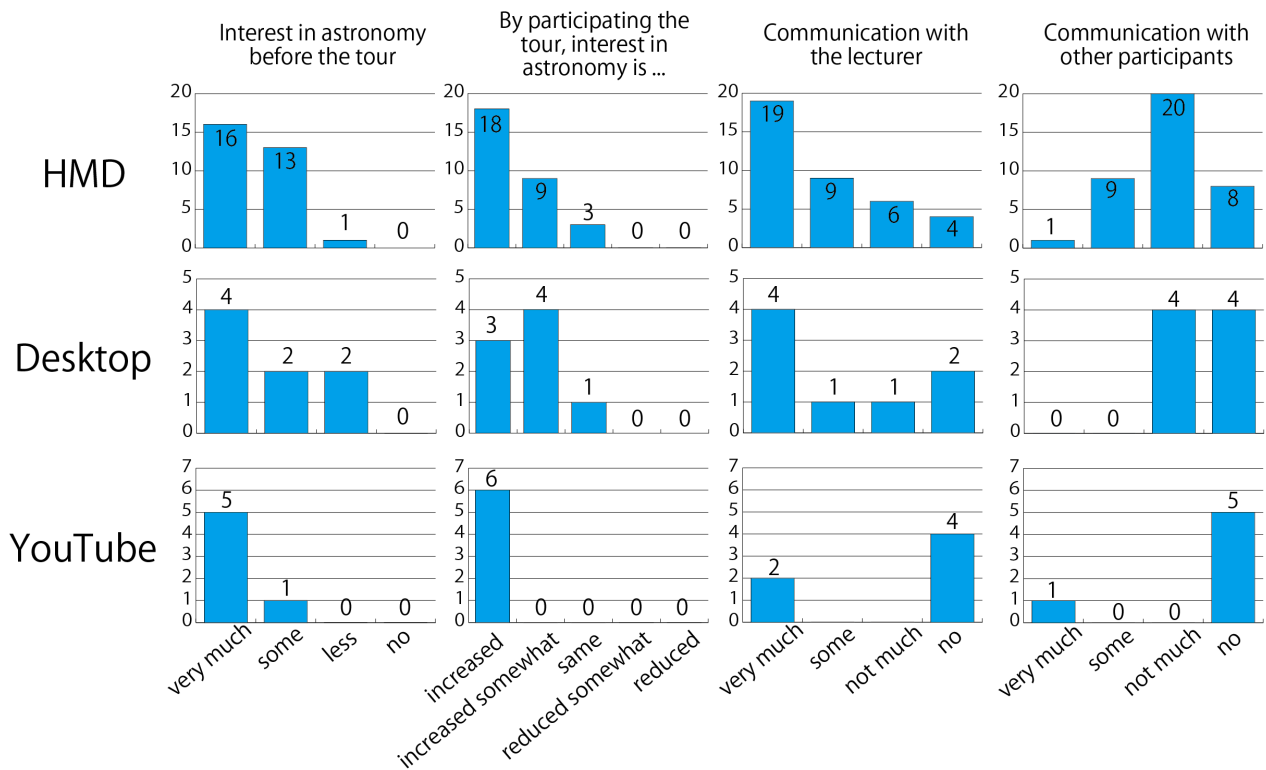

*Figure 7.* Summary of survey responses compiled by participating environment. Credit: Masaaki Hiramatsu

Phrases like "my own experience," "right in front of me," "right next to the telescope," and "experience as if I was actually there" appeared in the responses shows clear evidences that the participants felt a sense of reality in the CG environment and had positive impressions to understand the world and ALMA itself. Only two people (including the guide) in the tour had actually visited the ALMA Observatory, but both of them felt that the VR experience was very similar to an on-site visit. An interesting point is that the responses from those who watched the streaming on YouTube included words like "realistic sensation", but words based on physical sensations like "in front of me" did not appear. Although the participants via YouTube had a vivid impression through the tour because of the high level of reproduction of the ALMA Observatory in VRChat, the difference in the responses indicates that experience with HMD provides a good sense of immersiveness. The immersiveness of VR experiences using HMDs is also reported in Kersting et al. (2021). In their VR experience on the subject of gravitational wave astronomy, most of the participants described their immersive engagement as a positive experience.

In addition, the lecturer felt as if he was actually guiding the participants in the real observatory because he could feel the participants' "eyes" on him even though they were in avatars, and he also could hear the participants voice their impressions. Overall, for both the guide and the participants, the VRChat tour were felt similar to the actual on-site observatory tour. The ALMA Observatory is far away from most of the cities and it is difficult to visit. The pandemic is making it even more difficult, but the VRChat tour is a valuable way to experience the atmosphere in which cutting-edge astronomy is being performed.

As for the second point, six respondents pointed out the closeness to the lecturer. A representative answer was "it was very easy to ask questions because I was physically and psychologically close to the lecturer." Here we have two non-exclusive possible reasons why the psychological distance between the participant and the guide becomes closer in a guided tour in VR space. One is that the participants and the guide are spatially close to each other. In the tour participants could "stand" next to the lecturer through avatars. This is in contrast to typical face-to-face lectures where there is a distinct difference in locations between the lecturer and the audience. The other is that the use of friendly-looking avatars by both lecturers and participants increased anonymity and reduced the authority of the lecturer, creating an atmosphere where it was easy to talk to the lecturer. One respondent clearly stated that "participating with an avatar is fun and friendly."

It is interesting to discuss whether this closeness is characteristic of VR. Even in face-to-face activities, it is possible to minimize the physical distance between the lecturer and the audience by mixing the lecturer into the audience rather than having the lecturer stand behind the podium, and this may also contribute to reducing the psychological distance. In an in-person event, however, it is difficult to reduce the authoritative nature of the appearance of the lecturer, since the appearance of the lecturer cannot be changed. Nevertheless, wearing familiar costumes in the real world could also create a sense of discomfort because it is unnatural that only the lecturer has a different appearance than usual. On the other hand, in the VR space, since everyone participates using avatars, it is possible to reduce authority without creating a sense of discomfort.

## Difference between Conventional Virtual Events and VRChat

We also asked the participants about the difference between conventional online talk events and the tour in VRChat. Many participants expressed similar sentiments to those described in "Difference between Face-to-Face Events and VRChat". Notable points are:

1. A higher-quality audio-visual experience,
2. Participants can sense the responses of other participants.

As for the first point, VRChat provides an immersive experience in which one can get a 3D sense of place both visually and aurally. It is as if they were actually there. Some selected responses are as follows.

- "I enjoyed the realistic feeling of being able to see the antennas from a 360-degree direction, which is not possible with photographs."
- "It was easier to see and hear than a lecture through a two-dimensional screen."
- "The image quality of VR tours is much better than normal online lectures. The user experience is much better than Zoom lectures, and there are no distractions, so I was able to concentrate more on the lecture."

The immersive feeling cannot be obtained in normal 2-D virtual events. In terms of the comfort level, the participation of many people may overload the system and it is difficult to have hundreds of visitors at once in a VR environment. Proper handling of the total number of participants and data sizes of the avatars and the world provides a satisfying environment for virtual observatory tours.

Regarding the second point, one participant described that "Other participants' reactions were conveyed through their voices and gestures, which increased the sense of presence and live performance, and the lecturer seemed to be comfortable speaking." In virtual events on YouTube, participants can post their impressions in the text chat. However, compared to written communication, voice and gestures can achieve more direct communication, even if the participants are in avatars. Also, a respondent pointed out the possibility that VR would allow participants to talk with each other before and after the tour itself. This again shows the similarity between observatory visits in VR and actual visits.

**Other Responses**
In the free-write sections of the questionnaire, there were several comments pointing out the compatibility between VR and science communication. Notable responses are:
- "I was very impressed by seeing the details of ALMA, which we can't actually get close to, with explanations by the lecturer who actually works for it. The lecturer's explanation about the coldness and shortness of breath on site made me experience a more realistic atmosphere of the site, which complemented the VR experience."
- "It was an experience that made it stand out from other online talks. Although it will be very difficult to construct worlds and reproduce items one by one, it will be a great pleasure for the learners if cutting-edge researchers (in other research fields) give lectures in the same way, and it would also be very rewarding activity for lecturers."
- "I'm really happy to discover for the first time an event like this! It's a really good experience for me, I'm really happy to see some professionals join an event like this!"

In addition, we asked what the participants discovered during the tour. The purpose of this question was to understand how much the participants understood the guide's lecture and the tour, and what caught their attention. The answers ranged from the differences in the ALMA antennas developed by different countries, the meaning of their differences in shapes, and detailed information about telescope operations. The keywords we found in the responses are "realistic" and "detail". This showed that the various information disclosed by the guide during the tour was very well communicated to the participants, aided by the realism of the reproduced observatory in the VR world.

**Conclusions**
The virtual ALMA Observatory tour in VRChat was a very effective and fruitful experiment to overcome the difficulties of visiting remote astronomy observatories, especially under the global pandemic. The responses by the participants show the great potential of VR and also provide important insights into how to organise effective virtual tours. Important points we found are as follows.
- The quality of the virtual model of the observatory is essential. Recreating every detail increases the reality felt by the participants and the impression on the participants becomes stronger.
- It is important to cooperate with those who have high modelling skills and who can explain them properly. In this case, the collaboration between VSP and the NAOJ ALMA staff member worked very well and was key to the successful organisation of the tour.
- Even in the VR environment, sharing the same space enhances communications between the lecturer and participants, and among participants. The feeling was more similar to a face-to-face tour rather than a conventional virtual talk on YouTube. The lecturer has given many public lectures over YouTube, but he has not been able to go into much depth on topics because he could not see the audience's reactions and measure their level of understanding. On the other hand, with VRChat, even though the participants were in avatars and the guide could not see the faces directly,

- he saw that the participants were looking around the antennas and heard their murmuring their thoughts and interjections. Sensing these reactions he was able to confirm that the audience kept interest in his talk and provide further explanations at an appropriate level.
- Once the guide is sure that the participants are receptive to technical explanations, the guide can take full advantage of the detailed reproduction of the antennas in front of the participants to explain the technical details, which will be easier to understand than in other virtual events and also leave a strong impression.
- The technical difficulties for virtual tours using the VRChat world is the limitation of the number of participants in one event. In fact, there were several people who could not enter the world by the time due to the heavy access a few minutes before the start. One of the workarounds is to extend the waiting time for entering the world.
- Some participants suffered from technical glitches of their VR instruments and had difficulties in entering the world or showing the proper avatars. A few VSP members provided technical support during the tour. The VR environment is currently in the process of spreading, and there is a possibility that some participants may be unfamiliar with its operation and problems. When possible, it would be desirable to have someone who is familiar with the VR environment, like the VSP members this time, as support members.

The findings from our activity align well with those by Kersting et al. (2021). Despite the difference of the VR contents and the format of the experience, the importance of high-quality visualisation, interaction between professionals and participants, and collaboration is common to the two for realizing engaging and effective experience in science outreach.

Even after the pandemic, virtual tours in VR worlds will be very useful for the public to gain a deeper understanding of inaccessible astronomical facilities. On the other hand, it should be noted that the number of people who can use HMDs to participate in immersive VR events is still limited compared to the number of people who can browse YouTube, etc, as can be seen from the number of active users of VRChat. The VR equipment is rather expensive, and not many people may buy it just for scientific events. For the time being, until VR becomes more widespread, streaming the VR events to YouTube is one way to provide 3D virtual tours to a wider audience, even if the sense of immersion is somewhat lost.

VSP has made other telescope models such as NAOJ's Subaru Telescope and the Thirty Meter Telescope, as well as several scale models of space telescopes and space probes including the Hubble Space Telescope and the Hayabusa2. With the collaboration between astronomers/ communicators and someone with sufficient knowledge of Blender and Unity, other telescope models can also be built in the VRChat. A variety of effective virtual tours could be organised; possible programs are to compare different telescope models to understand the characteristics and importance of multi-wavelength astronomy, and to focus on the cutting-edge technologies that drive modern astronomy. It is impossible to actually visit the telescopes in remote sites instantaneously, however, we can do it in the VR world. We are looking forward to seeing many virtual tours providing a fun, vivid experience to visitors

## Acknowledgements
The authors thank the Kadinche Corporation

(https://www.kadinche.com/) for their technical support with the VR environment.

ALMA is a partnership of ESO (representing its member states), NSF (USA) and NINS (Japan), together with NRC (Canada), NSC and ASIAA (Taiwan), and KASI (South Korea), in cooperation with the Republic of Chile. The Joint ALMA Observatory is operated by ESO, AUI/NRAO and NAOJ.